\documentclass[prl,twocolumn,showpacs]{revtex4-1}
\usepackage{graphicx}
\usepackage{dcolumn}
\usepackage{bm}
\usepackage{amssymb}
\usepackage{amsmath}
\usepackage{epsfig}
\usepackage{epstopdf}

\begin{document}

\title{Molecular dissociation in presence of catalysts: Interpreting bond
breaking as a quantum dynamical phase transition}
\author{A. Ruderman$^{1,2}$, A. D. Dente$^{3}$, E. Santos$^{1,2}$, and H. M.
Pastawski$^{1}$}

\affiliation{$^{1}$Instituto de F\'{i}sica Enrique Gaviola (CONICET)-UNC and
Facultad de Matem\'{a}tica, Astronom\'{i}a y F\'{i}sica, Universidad
Nacional de C\'{o}rdoba, 5000, C\'{o}rdoba, Argentina}

\affiliation{$^{2}$Institute of Theoretical Chemistry, Ulm University,
D-89069 Ulm, Germany}

\affiliation{$^{3}$INVAP S.E., 8403, San Carlos de Bariloche, Argentina}

\date{\today}

\begin{abstract}
In this work we show that the molecular chemical bond formation and dissociation in
presence of the $d$-band of a metal catalyst can be described as a Quantum
Dynamical Phase Transition (QDPT). This agree with DFT calculations that predict sudden jumps in some observables as the  molecule breaks. According to our model this
phenomenon emerges because the catalyst provides for a non-Hermitian
Hamiltonian. We show that when the molecule approaches the surface, as
occurs in the Heyrovsky reaction of H$_{2}$, the bonding H$_{2}$ orbital has
a smooth crossover into a bonding molecular orbital built with the closest H
orbital and the surface metal $d$-states. The same occurs for the antibonding state. 
Meanwhile, two resonances appear within the
continuous spectrum of the $d$-band which are associated with bonding and
antibonding orbitals between the furthest H atom and the $d$-states at the
second metallic layer. These move towards the band center where they
collapse into a pure metallic resonance and an almost isolated H orbital.
This phenomenon constitutes a striking example of the
non-trivial physics enabled when one deals with non-Hermitian Hamiltonian
beyond the usual wide band approximation.
\end{abstract}

\maketitle

\section{Introduction}

When do two individual atoms become a molecule, or vice versa? This question
is a fundamental one as much for Chemistry as for Physics. Certainly, it
should involve a sort of discontinuity as in phase transition. In this
context, P. W. Anderson hinted, in his well known article \textquotedblleft
More is different\textquotedblright\ \cite{anderson1972more}, that a
condition for a phase transition is the presence of infinite degrees of
freedom. Sometimes, these are provided by the environment \cite%
{Pastawski2007278}.

Many results in DFT calculations show an abrupt change in chemical bonds as
the molecules approach to the surface of a catalyst and the interaction with
the metal increases. In this context, a paradigmatic example of molecular
formation and dissociation is the Heyrovsky reaction \cite{heyrovsky}, one
of the steps of the Hydrogen evolution reaction at metallic electrodes:
after the adsorption of a Hydrogen atom at the surface, a second proton
approaches and an electron is transferred from the metal. It is in this last
step, when the Hydrogen molecule is formed and a discontinuity is hinted by
DFT calculations as a jump in energy and spin polarization. This occurs at a
critical distance of the farthest Hydrogen \cite%
{BONDBREACK-Santos-2011-Diatomic-molecules}.

In Quantum Mechanics, a phase transition is recognized as a non-analytic
behavior of an observable, typically the ground state energy, as a function
of a control parameter. This phenomenon, absent in a few atoms system, is an
emergent property of the thermodynamic limit, i.e. when $N$, the number of
atoms or degrees of freedom, tends to infinity. \cite{sachdev2007quantum,
RedutionismBerry}. Such a limit is used to get the Fermi Golden Rule (FGR).
There, one assigns an infinitesimal imaginary part $-\mathrm{i}\eta $ to
each of the $N$ involved energies and $\eta $ is made zero only after taking 
$N\rightarrow \infty $ \cite{Pastawski2007278}. Thus, an initial energy $%
E_{0}$ acquires a finite energy uncertainty $\Gamma $ \ associated with a
decay rate $2\Gamma /\hbar $. A straightforward way to account for this
decay is to introduce an effective non-Hermitian Hamiltonian where some
energies acquire an imaginary component, e.g. $E_{0}\rightarrow E_{0}-%
\mathrm{i}\Gamma $. While this procedure dates back to E. Majorana \cite%
{majorana}, its deep physical implications only recently have become
evident. This paper seeks to rationalize the discontinuities found in DFT
calculations at the light of the Quantum Dynamical Phase Transition (QDPT)
concept \cite{Alvarez-LevsteinJCP2006}.

A clear experimental evidence of a dynamical transition showed up in NMR
experiments in a 2(CH)$_{5}$Fe crystal. There, the nuclear spins of the rare 
$^{13}$C-$^{1}$H dimers can perform Rabi oscillations. Beyond some critical
crystal orientation, these spins are seen to abruptly decouple. This occurs
when the\ interaction between the $^{1}$H spin and the rest of the crystal
becomes stronger than the $^{13}$C-$^{1}$H one \cite{Alvarez-LevsteinJCP2006}%
. This phenomenon is a QDPT and can also be interpreted as a particular case
of the superrandiance phenomenon predicted by Dicke \cite%
{Gross,Celardo,Liu,Hepp,Dicke}.

As a preliminary idea, let us explain how a tight-binding model for a
homonuclear diatomic molecule $A-B$ can show a non-analytical discontinuity.
As in the spin case discussed above, this simple model would display the
concepts relevant for the main discussion. Let $\delta E_{0}=2V_{AB}$ be the
usual bonding-antibonding splitting. If atom $B$, has an interaction with a
metallic band of width $W\gg \delta E_{0}$ (wide band approximation) its
energy acquires a finite lifetime $\hbar /2\Gamma $ due to the tunneling
into the metal ($\Gamma /2\ll W$). This results in the effective
non-Hermitian molecular Hamiltonian \cite{majorana},

\begin{equation}
\mathbb{H}_{\mathrm{eff.}}\mathbb{=}\left[ 
\begin{array}{cc}
E_{A} & -V_{AB} \\ 
-V_{AB} & E_{B}-\mathrm{i}\Gamma%
\end{array}%
\right] .
\end{equation}

The eigenenergies are now complex numbers. For small $\Gamma ,$ the
difference between their real parts is the splitting between the bonding and
antibonding molecular levels, $\delta E=\sqrt{\left[ \delta E_{0}\right]
^{2}-\Gamma ^{2}}.$ This splitting is now weakened by the interaction with
the substrate. Their corresponding imaginary parts $\Gamma /2$ are
identical. However, $\delta E$ has a non-analytic collapse when $\Gamma $\
reaches the critical value $\Gamma _{c}=\delta E_{0}.$ Beyond this
Exceptional Point (EP) \cite{Berry, Rotter1, Moiseyev, Rotter2}, the real
parts become degenerate and the imaginary parts bifurcate. For big $\Gamma $
values, the eigenvalue associated to $B$ has an imaginary part $\Gamma $ and
thus remains strongly mixed with the substrate. The other one, associated to 
$A$, has an uncertainty proportional to $\left\vert 2V_{AB}\right\vert ^{2}$/%
$\Gamma $, indicating an almost isolated orbital \cite{majorana}. A similar
phenomenon underlays the Quantum Zeno Effect \cite{Pascazio}, i.e. when
three orbitals interact, a strong interaction between two of them weakens
their interaction with the third. The detailed analytical and numerical
solution of the above model was discussed in great detail in the context of
QDPT by Dente et al. \cite{Dente}. However, while very appealing for its
simplicity, \textit{this picture can not be directly applied to a typical
metallic catalyst} because it is in a very different physical regime. Indeed,
 the weak interaction with the wide $sp$-band does afford for a
relevant role in the molecular dissociation. Thus, our attention should turn
to the strong interaction of the $d$-band with the molecule \cite{Norskov},
which has long been recognized as responsible for catalysis \cite%
{NewnsHydrogen-and-dband}. Yet the $d$-band\textit{\ }has a width $W_{d}$
smaller than the molecular level splitting $\delta E_{0}$, which prevents
using the wide band approximation. Thus, we are back with the question of 
\textit{which is the magnitude that shows a non-analyticity that could be
associated with a bond breaking in presence of a }$d$\textit{-band?}

In this paper, we answer this question by showing that an actual analytical
discontinuity appears if one includes a description of the metallic $d$-band
with the right degree of detail. This requires to choose an appropriate
combination between the molecular levels and the different metalic layers.
Furthermore, the Anderson-Newns theory of adsorption \cite%
{AndersonEli,NewnsHydrogen-and-dband} is needed to describe molecular dissociation
and electrocatalysis along with the
inclusion of further works that extended this theory \cite{Santos4}. In our
terms, while the molecule approaches to the surface, the farther $A$ atom
experiences a resonant through-bond coupling \cite{Levstein-Damato} with the
2nd layer of the metal. This interaction, mediated by the $B$ atom and the
first surface layer, manifests as two resonances inside the $d$-band. The
transition occurs when these resonances collapse at the center of the $d$%
-band, releasing the $A$ atom. Meanwhile, the $B$ atom hybridization with $%
A $ is swapped into a $B$-metal bonding.

\section{A model for molecule dissociation}

We consider the Heyrovsky reaction, i.e. a H$_{2}$ molecule approaching
perpendicularly to the metal surface. The molecule Hamiltonian is $\hat{H}%
_{S}=E_{A}\left\vert A\right\rangle \left\langle A\right\vert
+E_{B}\left\vert B\right\rangle \left\langle B\right\vert -V_{AB}\left(
\left\vert A\right\rangle \left\langle B\right\vert +\left\vert
B\right\rangle \left\langle A\right\vert \right) $. The degeneracy of the
atomic energies $E_{A}$ and $E_{B}$ is broken by the bonding interaction $%
V_{AB}$. According to the standard wisdom \cite{Hoffman} the $d_{z^{2}}$
(top) or combination of the $d_{xz}$ and $d_{yz}$ (for hollow sites) are the
only $d$-orbitals with a finite overlap with the molecule. The orbital $%
\left\vert B\right\rangle $ interacts with the closest $d$-orbital
combination, say $\left\vert 1\right\rangle $, pointing along the connecting
path through the binding energy $V_{0}$, $\hat{V}_{SM}=-V_{0}\left(
\left\vert B\right\rangle \left\langle 1\right\vert +\left\vert
1\right\rangle \left\langle B\right\vert \right) $. The interaction energy $%
V_{0}$ is roughly an exponential function of the molecule-substrate
distance. Therefore, the complete Hamiltonian becomes $\hat{H}=\hat{H}_{S}+%
\hat{H}_{M}+\hat{V}_{SM}$.

\begin{figure}[h]
\includegraphics[width=1.0\linewidth]{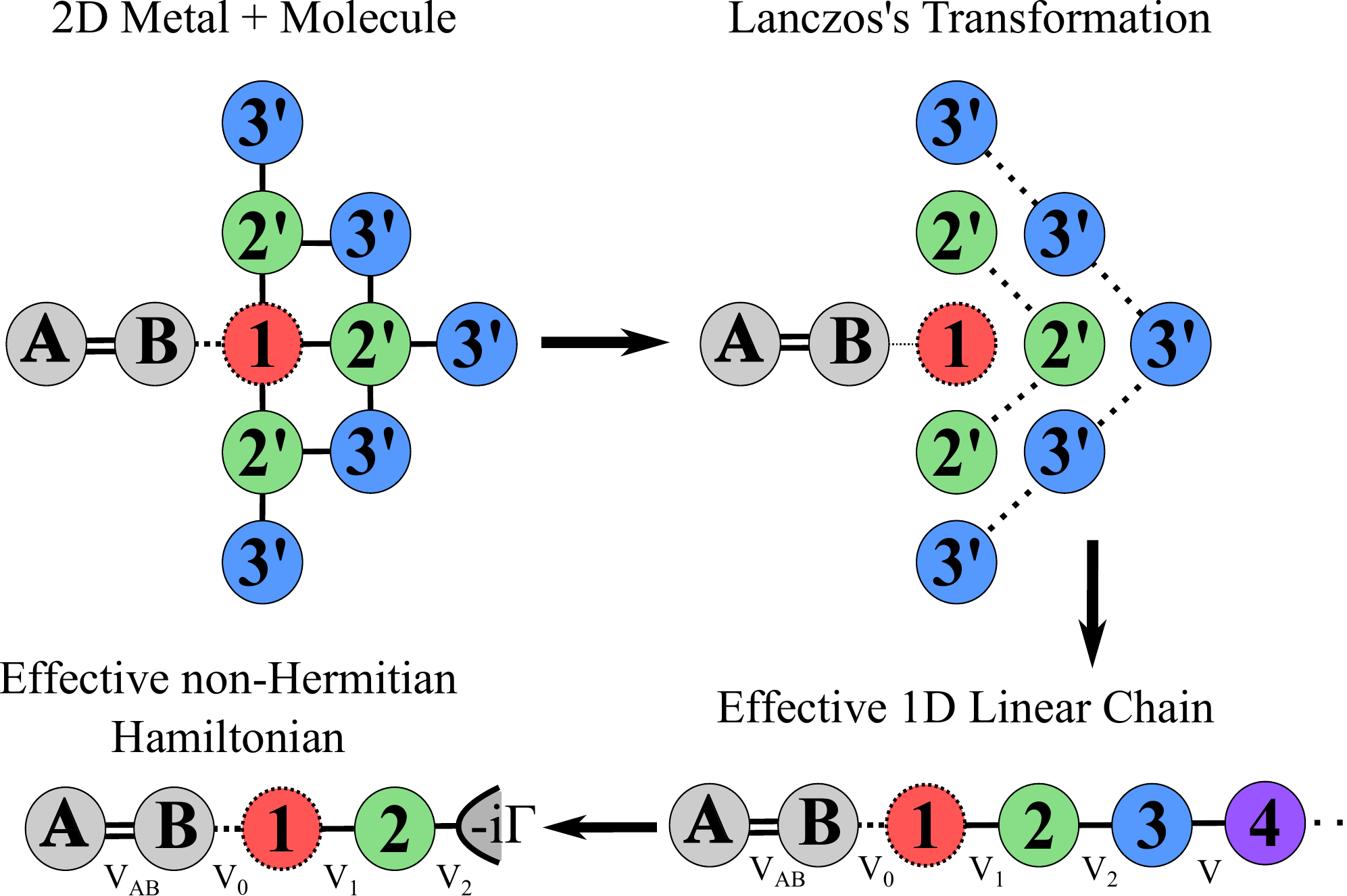}
\caption{Effective non-Hermitian Hamiltonian: diatomic A-B molecule (in
gray) in a configuration perpendicular to a $2D$ metal. The Lanczos unitary
transformation combines orbitals at the same distance. The resulting
tridiagonal Hamiltonian can be represented as a linear chain. A decimation
procedure leads to a $4\times 4$ non-Hermitian Hamiltonian. The same
procedure can be applied to the $3D$ metal. Dot 1 represents a single atom for $top$ interaction
or a combination of them for $hollow$.}
\label{lanczos}
\end{figure}

Let us now focus on $\hat{H}_{M}$ which describes the $d$-band. Since the
Newns pioneering work \cite{NewnsHydrogen-and-dband}, it is usually assumed
that a semi-elliptical shape is a good approximation for the Local Density of
States (LDoS):

\begin{equation}
N_{d}\left( \varepsilon \right) =\frac{1}{\pi W_{d}}\sqrt{%
W_{d}^{2}-4\varepsilon ^{2}}\times \Theta \left[ W_{d}-2\varepsilon \right]
\times \Theta \left[ 2\varepsilon -W_{d}\right] ,  \label{e4}
\end{equation}%
where $\Theta \left[ x\right] $ is the Heaviside function. Indeed, the
validity of this proposal for an actual metal, can be visualized through the
Lanczos transformation\cite{haydock1972electronic}. This is a unitary
transformation which maps the actual $3D$ substrate into an equivalent $1D$
linear chain. Fig. \ref{lanczos} represents this procedure for a $2D$ case.
Starting from the $\left\vert 1\right\rangle $ surface $d$-orbital, a
sequence of \textquotedblleft collective orbitals\textquotedblright\ is
build up mainly from the atomic orbitals at layers progressively distant
from the original surface orbital. Regardless of the precise original
Hamiltonian $\hat{H}_{M}$, in the Lanczos basis the metal Hamiltonian $\hat{H%
}_{M}^{L}$ has identical \textquotedblleft site\textquotedblright\ energies $%
E_{n}\equiv E_{d}$ and only nearest neighbor interactions $V_{n}$'s that
account for the coupling among the Lanczos collective states \cite%
{haydock1972electronic}. Their rapid convergence to $V_{\infty }\equiv
V=W_{d}/4$ justifies the Newns proposal of a homogeneous linear chain to
describe the actual metal. The kets $\left\vert n\right\rangle $ are the
collective $d$-orbitals, now\ the $n$th site of the metallic chain. Since in
the surface of the metal the coupling between layers is weaker than in the
bulk we assume $\left\vert V_{1}\right\vert \lesssim \left\vert
V_{2}\right\vert \lesssim \left\vert V\right\vert $ and we set $V_{n}\equiv
V $ $\ $for $n\geq 3.$

\begin{equation}
\hat{H}_{M}^{L}={\displaystyle\sum\limits_{n=1}^{\infty }}E_{n}\left\vert
n\right\rangle \left\langle n\right\vert +{\displaystyle\sum\limits_{n=1}^{%
\infty }-}V_{n}\left( \left\vert n\right\rangle \left\langle n+1\right\vert
+\left\vert n+1\right\rangle \left\langle n\right\vert \right) .  \label{e5}
\end{equation}

With the purpose to define an optimal configuration for the molecular
dissociation we will base our model in the Anderson-Newns theory and set up
the Fermi energy level at $E_{d},$ the center of the $d$-band, thus $%
E_{A}=E_{B}=E_{d}$. Thus the molecular bonding\ $V_{AB}$ yields a symmetric
splitting around the center of the band. Additionally, we will set the
coupling elements $V_{1}/V=0.8$, $V_{2}/V=0.9$, and $V_{AB}/V=2.5$ which
results fairly representative of various situations \cite{Santos2011314}.

At this point it is necessary to list some of the main approximations
implied by this model: 1) The fixed value of $V_{AB}$ neglects the variation
of the distance between the atoms $A$ and $B$. This does not affect our main
results since the molecule breaking can be viewed as a competence among
interactions 2) As usual the atoms in the metal are considered fixed in the
whole problem. However, variations are minor in the Lanczos approach 3) We
assume a null coupling between the metal and $A$ the furthest atom. A
residual exponentially small interaction would be completely masked by the
through-bond interaction.

\section{Molecule dissociation as a spectral bifurcation}

The solution of a linear chain model is better expressed in terms of the Retarded
Green's function matrix $\mathbb{G}=(\varepsilon \mathbb{I}-\mathbb{H})^{-1}$%
, whose divergences occur at the Hamiltonian eigenstates. For example, in
absence of the metal we have:

\begin{equation}
G_{AA}^{(S)}(\varepsilon)=\frac{1}{\varepsilon-E_{A}-\Sigma_{A}(\varepsilon )%
},\text{ where~~}\Sigma_{A}(\varepsilon)=\frac{\left\vert V_{AB}\right\vert
^{2}}{\varepsilon-E_{B}}.  \label{e7}
\end{equation}

Clearly, $E_{A}$, the isolated $A$ atom energy, is modified by the presence
of $B$ through the self-energy $\Sigma _{A}(\varepsilon ),$ a real function
accounting for the bonding and providing for the exact molecular
eigenstates. As discussed before, one uses $\tilde{E}_{k}=E_{k}-\mathrm{i}%
\eta $ (for $k=A,B,n$). This regularization energy $\eta $ facilitates the
study of the spectral density and the retarded charge dynamics, and whose
physical origin can be traced back to small \ \textquotedblleft
environmental interactions\textquotedblright , as with the infinite $sp$%
-band states\cite{CattenaBustosPRB2010}. When the molecule is coupled with
the metal in presence of the infinitesimal environment, one gets the
retarded Green's function \cite{pastawski-medina}:

{\small {%
\begin{equation}
G_{AA}(\varepsilon )=\dfrac{1}{\varepsilon -\tilde{E}_{A}-\dfrac{|V_{AB}|^{2}%
}{\varepsilon -\tilde{E}_{B}-\dfrac{|V_{0}|^{2}}{\varepsilon -\tilde{E}_{1}-%
\dfrac{|V_{1}|^{2}}{\varepsilon -\tilde{E}_{2}-\dfrac{|V_{2}|^{2}}{|V|^{2}}%
\Sigma (\varepsilon )}}}}.  \label{e13}
\end{equation}%
} }Here, $\Sigma (\varepsilon )$ is the self-energy correction describing
the bulk of the metal $d$-band in the Lanczos representation:

\begin{equation}
\Sigma (\varepsilon )=\frac{\left\vert V\right\vert ^{2}}{\varepsilon -%
\tilde{E}_{d}-\Sigma (\varepsilon )}=\Delta (\varepsilon )-\mathrm{i}\Gamma
(\varepsilon )  \label{e9}
\end{equation}
From now on we set $E_{d}=0,$ thus, the solution of Eq. \ref{e9} results 
\cite{Cattena--FernandezJPCM}:

\begin{equation}
\Sigma (\varepsilon )=\dfrac{\varepsilon +\mathrm{i}\eta }{2}%
-sgn(\varepsilon )\times \left( \sqrt{\dfrac{r+x}{2}}+\mathrm{i}sgn(y)\times 
\sqrt{\dfrac{r-x}{2}}\right) ,  \label{e10}
\end{equation}%
with $x=\dfrac{\varepsilon ^{2}-\eta ^{2}}{2}-V^{2}$, $y=\dfrac{\varepsilon
\eta }{2}$ and $r=\sqrt{x^{2}+y^{2}}$. For the effective site at the second
layer: 
\begin{equation}
G_{22}(\varepsilon )=\dfrac{1}{\varepsilon +\mathrm{i}\eta -\Sigma
_{2L}(\varepsilon )-\dfrac{|V_{2}|^{2}}{|V|^{2}}\Sigma (\varepsilon )},
\label{EqG22}
\end{equation}%
with 
\begin{equation}
\Sigma _{2L}(\varepsilon )=\dfrac{|V_{1}|^{2}}{\varepsilon +\mathrm{i}\eta -%
\dfrac{|V_{0}|^{2}}{\varepsilon +\mathrm{i}\eta -\dfrac{|V_{AB}|^{2}}{%
\varepsilon +\mathrm{i}\eta }}},  \label{EqSigma2L}
\end{equation}%
and similar equations for the other sites. In all these cases the LDoS is
obtained from the imaginary part of the Green's function $N_{i}\left(
\varepsilon \right) =-\frac{1}{\pi }\lim_{\eta \rightarrow 0}\mathrm{Im}%
\left[ G_{ii}\left( \varepsilon \right) \right] .$

The imaginary part of the self-energy accounts for the quantum diffusion of
the electrons in the metallic substrate. Notice that the imaginary parts of
the self-energy, $\Gamma (\varepsilon )$, and the continuum spectrum are
already consequences of taking the thermodynamic limit of infinitely many
sites in the chain. Otherwise $\Sigma (\varepsilon )$ would be a collection
of divergences at discrete eigenenergies as $\Sigma _{2L}(\varepsilon )$.
Within the $d$-band, the imaginary part survives the limit $\eta \rightarrow
0,$ indicating that each atomic orbitals merges into the metallic band \cite%
{anderson1978local}. However, the mere existence of $\Gamma \neq 0$ does not
warrant the QDPT. In this narrow band limit, a QDPT emerges as consequence
of the specific non-linear dependence of $\Gamma $ and $\Delta $ on $%
\varepsilon $ that accounts for the different metal layers.

Now, the important distinction respect to the introductory example is that
the self-energies acquire a non-linear dependence on $\varepsilon $ that
contains all the wealth of the molecule-catalyst interaction. Finding the
corresponding energy spectrum involves a 4th order polynomial on $%
\varepsilon $ with complex coefficients. A simple procedure is to find the
eight complex roots of $\left\vert 1/G_{AA}(\varepsilon )\right\vert ^{2}$.
Half of them are non-physical as they are divergences for $\left\vert
G_{AA}(\varepsilon )\right\vert ^{2}$ but not for $G_{AA}(\varepsilon )$.
Thus, we evaluate the poles numerically. Once we obtain the solutions, we
choose the physical ones, i.e. those whose imaginary component is negative,
i.e. poles of the retarded Green's function.

In Fig. \ref{f:reposeful} we show the real part of the Green's function poles.
There, we observe two energies outside the $d$-band, which for $V_{0}=0$
correspond to the bonding ($\left\vert A\right\rangle +\left\vert
B\right\rangle $)/$\sqrt{2}$ (shorted as $\left\vert AB\right\rangle $) and
antibonding ($\left\vert A\right\rangle -\left\vert B\right\rangle $)/$\sqrt{%
2}$ (shorted as $\left\vert \left( AB\right) ^{\ast }\right\rangle $)
localized states of the lonely molecule (i.e. H$_{2})$ at $\mp V_{AB}$. When 
$V_{0}$ increases strongly, e.g. for $V_{0}=3V$, these energies split
furthermore and become a bonding and anti-bonding states between $\left\vert
B\right\rangle $ and $\left\vert 1\right\rangle $ orbitals, $\left\vert
B1\right\rangle $ and $\left\vert \left( B1\right) ^{\ast }\right\rangle $,
respectively. As in the quantum Zeno effect, increasing this interaction
would dissociate the $A$ atom from the rest of the system \cite{Pascazio}.
This does not preclude a small amount of tunneling of $A$ into the
substrate's second layer passing through the orbitals $B$ and 1. For
intermediate $V_{0}$, this originates a through-bond interaction \cite%
{Levstein-Damato} that favors the formation of a bonding state between $%
\left\vert A\right\rangle $ and\ $\left\vert 2\right\rangle $ , as well as
an antibonding one($\left\vert A2\right\rangle $ and $\left\vert \left(
A2\right) ^{\ast }\right\rangle $ respectively). They are not localized
states but resonances, as the electrons can be exchanged with the bulk.

\begin{figure}[tph]
{\small \includegraphics[width=1.0\linewidth]{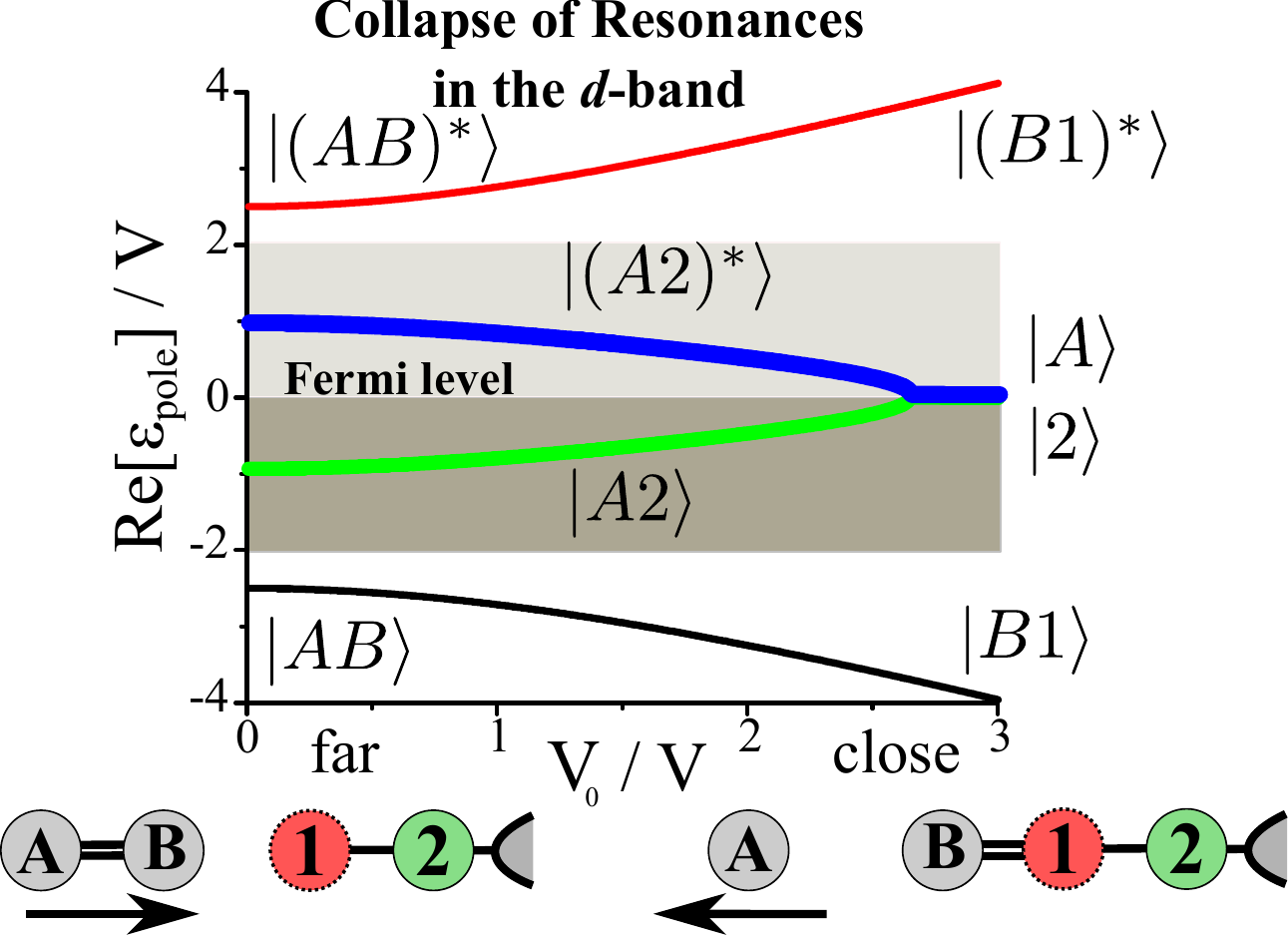} }
\caption{Collapse of resonances in the $d$-band. The real part of the poles
is plotted versus the molecule-substrate interaction,  V$_{0}$. Small V$_{0}$
values represent a far away molecule. For larger V$_{0}$'s the molecule is
close to the metal. The gray area represents the $d$-band, which goes from
-2V to 2V. Outside the band the bonding molecular orbital $\left\vert
AB\right\rangle $ smoothly becomes a bonding combination between the atom
and the metal $\left\vert B1\right\rangle $ (black curve). The same occurs
for the antibonding state $\left\vert (AB)^{\ast }\right\rangle $ that
becomes $\left\vert (B1)^{\ast }\right\rangle $ (red curve). Poles within
the d-band correspond to the bonding and antibonding resonances $\left\vert
A2\right\rangle $ and $\left\vert (A2)^{\ast }\right\rangle $ (green and
blue curves respectively), after the transition they become an almost
isolate $\left\vert A\right\rangle $ orbital and a $\left\vert
2\right\rangle $ orbital strongly coupled to the metal.}
\label{f:repolos}
\end{figure}

\begin{figure}[tph]
{\small \includegraphics[width=0.85\linewidth]{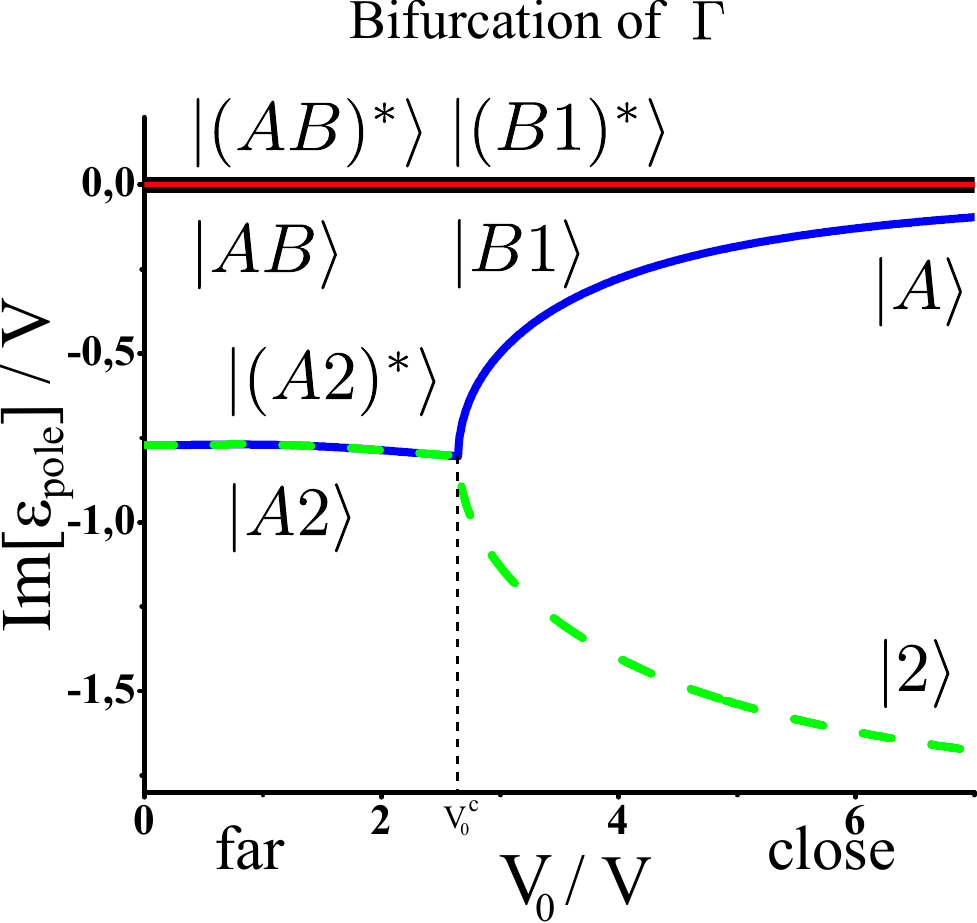} }
\caption{Bifurcation of Decay Rates. The imaginary part of the poles is plotted
versus the V$_{0}$ molecule-substrate interaction. Small V$_{0}$ values
represent a far away molecule. Black and red lines over the abscissa account
for the infinite lifetime of the localized states out of the band. At $%
V_{0}^{C}$ the imaginary part of the resonances (blue and green curves) show
a bifurcation. One branch accounts for a the increasingly long lived atomic
level, and the other branch describes the uncertainty of $\left\vert
2\right\rangle $ that transforms it into the metallic delocalized band.}
\label{f:impolos}
\end{figure}

The above mentioned resonances, appear as poles of the Green's function with
a finite imaginary part accounting for their coupling with the metal, Fig. %
\ref{f:impolos}. However, and this is the main point of this work, when $%
V_{0}$ increases and reaches $V_{0}^{C}$, a quantum phase transition occurs
and the state $\left\vert A\right\rangle $ becomes almost purely atomic.
Simultaneously, state $\left\vert 2\right\rangle $ recovers its purely
metallic nature. At this transition the bonding and antibonding resonances
(identified by the real parts of the poles) collapse into a degenerate
value. At this precise $V_{0}^{C}$ the pole's imaginary parts have a
non-analytic bifurcation into a decreasing part, which accounts for a long
lived atomic level, and an increasing uncertainty that represents the
metallic delocalization of $\left\vert 2\right\rangle $. This process can be
interpreted as a manifestation of the Quantum Zeno Effect meaning that when
the interaction $V_{0}$ between atom $B$ and the metal increases, the
interaction between $A$ and $B$ becomes weaker \cite{Pascazio,
Alvarez-LevsteinJCP2006}. In addition, the imaginary part of the states out
of the band remain zero for the whole range of $V_{0}$, due to the fact that
they are localized states with an infinite lifetime.

\begin{figure}[tbp]
{\small \includegraphics[width=1.0\linewidth]{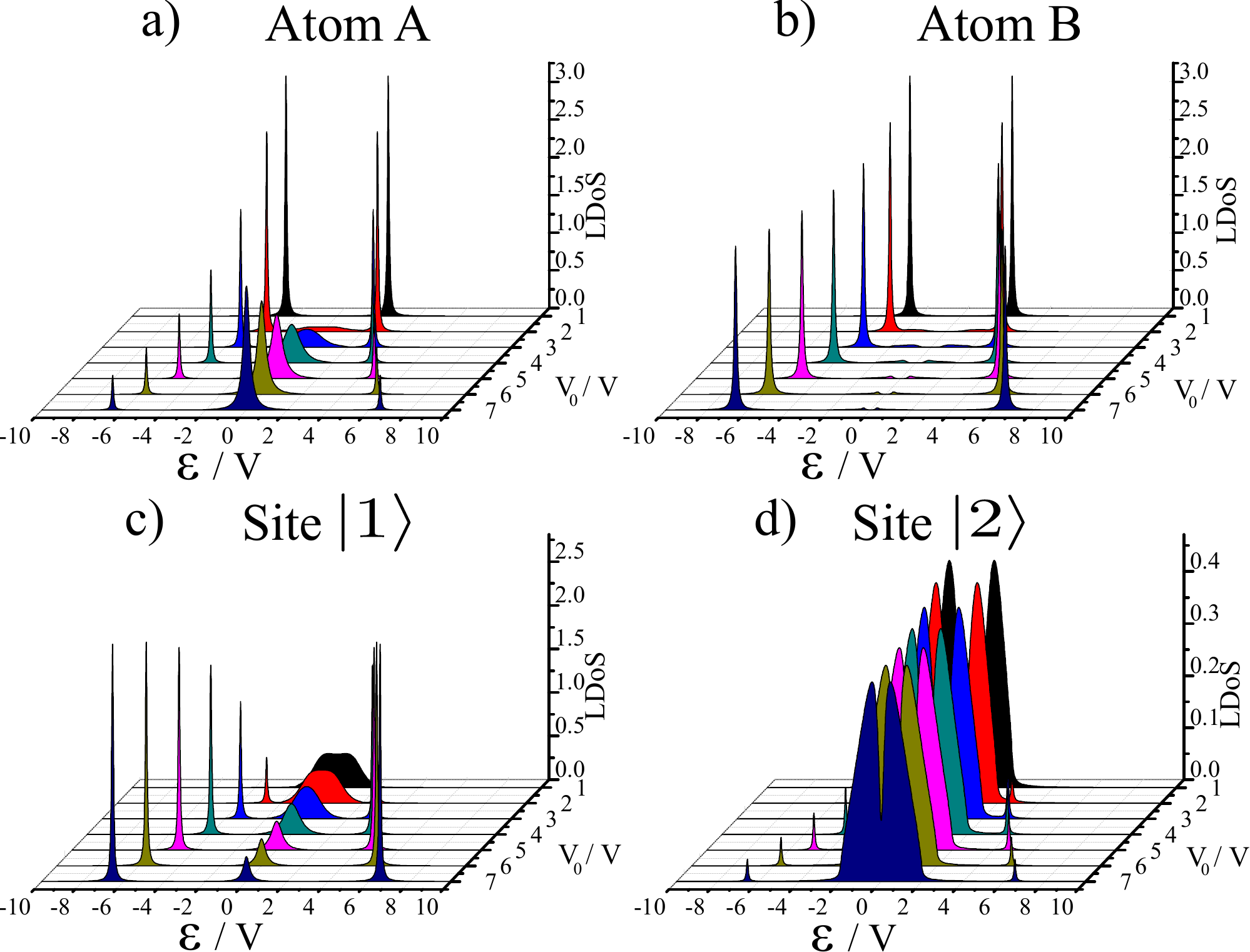} }
\caption{LDoS for different atoms and metallic sites. a) and b) $A$ and $%
B$ atomic orbitals, respectively. c) and d) $1$ and $2$ effective metal
orbitals, respectively. For atom $A$ we observe a decrease of the LDoS over
the energies outside the band and an increment of its participation on the $%
d $-band resonances. Atom $B$, instead, does not lose its participation over
the states outside the $d$-band as $V_{0}$ increases. The metallic orbital $%
\left\vert 1\right\rangle $ loses its participation inside the band and
gains presence over the localized states. The LDoS at orbital $\left\vert
2\right\rangle $ shows almost no participation on the localized states
outside the band. Consistently, the anti-resonance at the band centers
ensures no mixing with $\left\vert A\right\rangle $.}
\label{f:densidades}
\end{figure}

The detailed analysis of the spectral properties can be correlated with a
study of the LDoS at different orbitals. Fig. \ref{f:densidades} shows the
results from such evaluation. In a) and b), we can see that for long
distances ($V_{0}\simeq 0$) the LDoS at $\left\vert A\right\rangle $ and $%
\left\vert B\right\rangle $ show a dominant presence outside the $d$-band.
In contrast, c) and d) show that the sites $1$ and $2$ of the metal do not
have an appreciable participation at these energies. As the molecule
approaches the surface and $V_{0}$ increases, the LDoS at the orbital $%
\left\vert A\right\rangle ,$ shown in Fig. \ref{f:densidades} a), looses its
weight over the states outside the $d$-band. When $V_{0}\approx V$, we
observe the emergence of two broad resonances inside the band accounting for
the $\left\vert A2\right\rangle $ and $\left\vert \left( A2\right) ^{\ast
}\right\rangle $ orbitals. Close to the non-analyticity point $V_{0}^{C}$,
we observe that both resonances collapse into a single peak at $\varepsilon
=E_{A}=0$. This is precisely the energy of an electron at the isolated
orbital $\left\vert A\right\rangle $.

An interesting complementary behavior is observed on the LDoS at $\left\vert
2\right\rangle $. Fig. \ref{f:densidades} d) shows that, as $V_{0}$
increases, the two separate peaks, typical of a second layer \cite%
{schrieffer2008theory}, become close and almost collapse. They are still
separated by an anti-resonance \cite{Levstein-Damato}, i.e. a destructive
interference with $\left\vert A\right\rangle $ which manifest as pole for
Eq. \ref{EqSigma2L}.

In Fig. \ref{f:densidades} c) we observe that, as $V_{0}$ increases, the
first metal site starts loosing participation on the $d$-band.
Simultaneously, it increases its participation on the bonding and
antibonding states localized outside the $d$-band. Accordingly, the $%
\left\vert B\right\rangle $ orbital maintains a substantial weight in these
localized states (Fig. \ref{f:densidades} b) ). This accounts for the fact
that after the phase transition, when atom $A$ decouples from $B$, the
out-of-band localized states become the bonding and antibonding $\left\vert
B1\right\rangle \ $and $\left\vert \left( B1\right) ^{\ast }\right\rangle $.

\begin{figure}[tbh]
\center{\small {\includegraphics[width=1\linewidth]{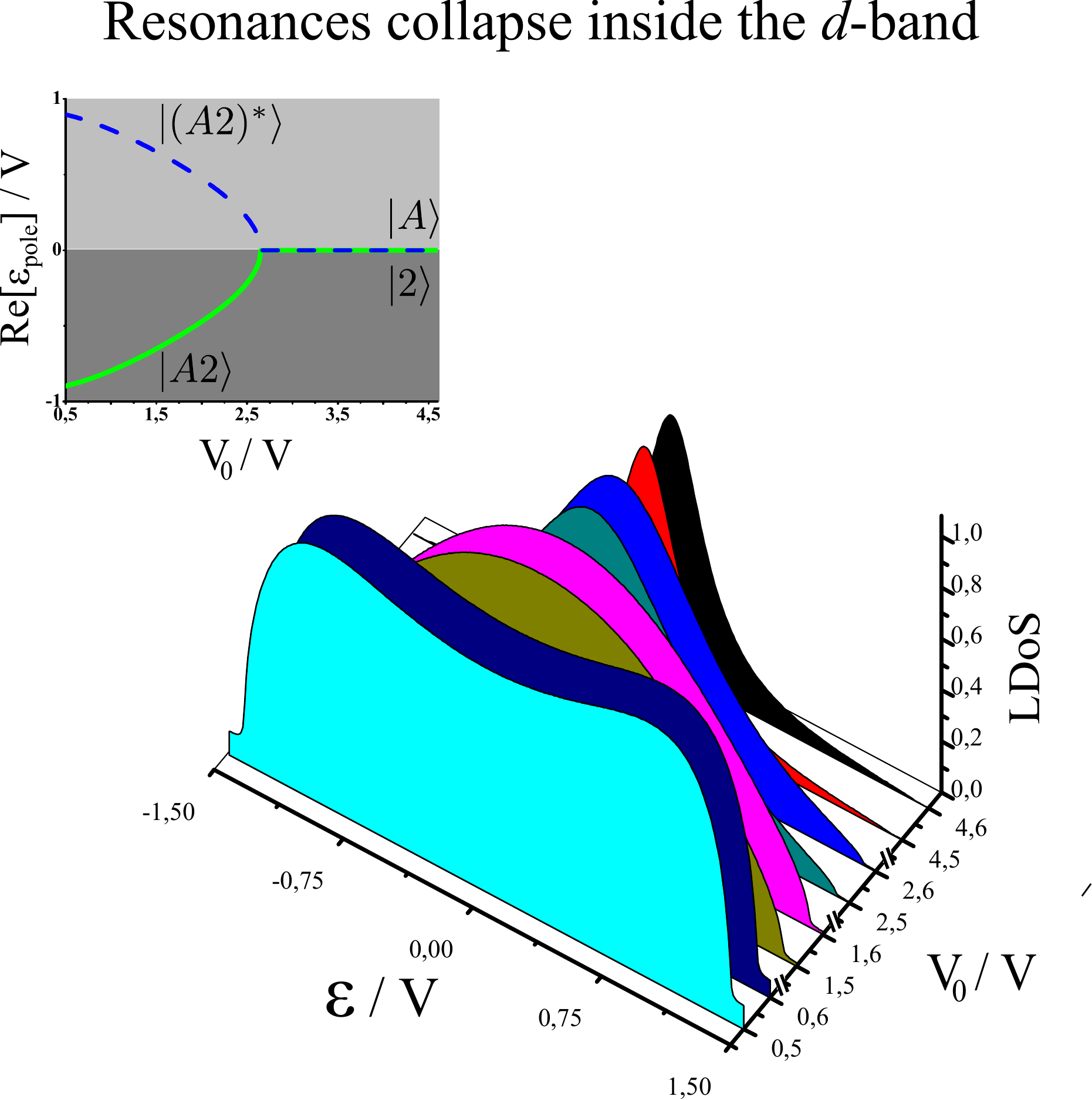} } }
\caption{ LDoS of $A$ inside the $d$-band normalized to the local maximum.
There we can see the two broad resonances collapse as $V_{0}$ reaches $%
V_{0}^{C}$. This collapse describes the Quantum Dynamical Phase Transition.
Furthermore, after $V_{0}^{C}$ the LDoS becomes narrower as $A$ becomes
isolated.}
\label{f:resonancias}
\end{figure}

One might wonder how these prediction match with those of more realistic DFT
calculation, such as the H$_{2}$ molecule interacting with a silver catalyst 
\cite{BONDBREACK-Santos-2011-Diatomic-molecules}. A first interesting
effect, reproduced by DFT calculations and observed in the tight binding
model, is the screening effect that the furthest atom suffers because of the
presence adsorbed one. This result is observed in our LDoS on the bonding
energies outside the $d$-band of Fig. \ref{f:densidades}. There, the
participation of the $\left\vert A\right\rangle $ orbital is smaller than
that of $\left\vert B\right\rangle $. However, the highly structured $d$%
-band masks the spectral branching that characterizes the QDPT evident in
our tight-binding model, Fig. \ref{f:resonancias}, after a renormalization
to the LDoS maximum. Nevertheless, in a DFT, the hidden spectral bifurcation
still constitutes the input for the full non-linear self-consistent
calculation. Thus it triggers the discontinuity on some of the observables
that show up. Indeed, while a Hydrogen atom approaches to another one
adsorbed at the silver surface, a sudden jump is observed in the total
energy of the system. At the same time, the adsorbed atom also jumps to a
position close to the approaching one, forming the H$_{2}$ molecule. Thus,
the Heyrovsky reaction shows discontinuities whose roots may be assigned to
the spectral discontinuities described by our QDPT model.

\section{Conclusions}

We analyzed the molecular dissociation in an Heterogeneous Catalysis process
under the framework of Quantum Dynamical Phase Transitions \cite%
{Alvarez-LevsteinJCP2006,Rotter1, Rotter2}. As hinted by P. W. Anderson \cite%
{anderson1978local}, the non-analyticity of the observables is an emergent
phenomenon enabled by an infinite number of environmental degrees of
freedom, in this case provided by the catalyst's $d$-band. We first observe
a smooth crossover of the localized bonding and antibonding molecular
states, which lie outside the narrow $d$-band, into a bonding and
antibonding combination between the closest atomic orbital and the first
layers of the metal $d$ orbitals. By reducing the LCAO model to a
non-Hermitian Hamiltonian where the imaginary parts have specific non-linear
dependence on energy, we show that this system undergoes a collapse of
resonances that provides the key to understand the dissociation phenomenon.
\ More specifically, each of the resonances is formed from the bonding and
antibonding interaction between the furthest atom and a combination of $d$
orbitals at the second layer of the metal. Before the molecule dissociation,
both resonances are equivalently broadened by the rest of the metal. However, due to
thw interaction with the surface they
merge into a collective metallic molecular orbital centered in the second
layer of the substrate and an isolated atomic orbital at the center of the $%
d $-band.

In summary, we show that molecular dissociation constitute a striking
example of the Quantum Dynamical Phase Transition, a simple but non-trivial
phenomenon that only could emerge because we dealt with the thermodynamic
limit through non-Hermitian Hamiltonians.

\section*{Acknowledgments}

We acknowledge the financial support from CONICET (PIP 112-201001-00411),
SeCyT-UNC, ANPCyT (PICT-2012-2324) and DFG (research network FOR1376). We
thank P. Serra and W. Schmickler for discussions and references.

\end{document}